

\let\oldeqno = \eqno
\def\eqno#1{\oldeqno{\hbox{#1}}}

\def\pmb#1{\setbox0=\hbox{$#1$}%
\kern-.025em\copy0\kern-\wd0
\kern.05em\copy0\kern-\wd0
\kern-.025em\raise.0433em\box0}

\def\pp{\partial}
\def\cl{{\cal L}}
\def\kappal{\epsilon}
\def\sqg{{\scriptstyle\sqrt{\mathstrut {\textstyle -g}}}}
\def\sg{{\scriptstyle\sqrt{\mathstrut {\textstyle g}}}}

\def\TC{\Theta\kern-.1em\lower0.6ex\hbox{${}_{C}$}}

\magnification=1200

\hsize=6.7truein \hoffset=-.1in
\vsize=9.2truein \voffset=-.1in

\tolerance 600

\font\eightrm = cmr8
\def\footnoterule{\kern-3pt \hrule width \hsize \kern2.6pt}
\pageno=0 \footline={\ifnum\pageno>0 \hss --\folio-- \hss \else\fi}
\baselineskip 12.75pt plus .5pt minus .5pt
\parskip=2pt

\centerline{\bf ENERGY-MOMENTUM CONSERVATION}
\smallskip
\centerline{{\bf IN GENERAL RELATIVITY}\footnote{*}
{\eightrm This work is supported in part by funds provided by
the U.~S.~Department of Energy (D.O.E.) under contract
\#DE-AC02-76ER03069 and in part by
the Fonds du 450e de l'Universit\'e de Lausanne (D.C.)}}
\vskip 24pt
\centerline{Dongsu Bak, D. Cangemi, and R. Jackiw}
\vskip 12pt
\centerline{\it Center for Theoretical Physics}
\centerline{\it Laboratory for Nuclear Science}
\centerline{\it and Department of Physics}
\centerline{\it Massachusetts Institute of Technology}
\centerline{\it Cambridge, Massachusetts~~02139~~~U.S.A.}
\vfill
\midinsert
\narrower
\centerline{\bf ABSTRACT}
\medskip
{ \baselineskip 15pt plus .5pt minus .5pt
We discuss general properties of the conservation law associated with a local
symmetry.  Using Noether's theorem and a generalized Belinfante symmetrization
procedure in 3+1 dimensions, a symmetric energy-momentum (pseudo) tensor for
the gravitational Einstein-Hilbert action is derived and discussed in
detail. In 2+1 dimensions, expressions are obtained for energy and angular
momentum arising in the $ISO(2,1)$ gauge theoretical formulation of Einstein
gravity.  In addition, an expression for energy in a gauge theoretical
formulation of the string-inspired 1+1 dimensional gravity is derived and
compared with the ADM definition of energy. \par}
\endinsert
\vfill
\centerline{Submitted to: {\it TBA}}
\vfill
\line{CTP \#2245 \hfil hep-th/9310025 \hfil October 1993}
\eject

\goodbreak\bigskip
\line{\bf I.~~INTRODUCTION \hfil}
\nobreak\medskip\nobreak\noindent
The definition of energy and momentum in general relativity has been under
investigation for a long time. The problem is to find an expression that is
physically meaningful and related to some form of continuity equation,
$$ \partial_\mu j^{\mu} = 0 \eqno{(1)} $$
which leads to a conserved quantity,
$$ Q = \int_{V}dV \, j^{0} \eqno{(2)} $$
provided $\int_{\partial\, V}dS^i \, j^{i}$ vanishes at infinity.
Therefore to insure conservation of $Q$, $j^{i}$ has to satisfy suitable
boundary conditions. In other words, to get a conserved quantity from a
continuity equation we always need to specify asymptotic behavior.

In field theory, conservation equations are usually related to invariance
properties of the action, in which case the conserved current is called a {\it
Noether current}.  The Einstein-Hilbert action is invariant under
diffeomorphisms, which are local transformations; more specifically, it is
invariant under Poincar\'e transformations, which comprise special
diffeomorphisms and can be viewed as ``global'' transformations.

In 3+1 dimensions, asymptotically Minkowski boundary conditions can be posed,
so that we can associate energy, momentum and angular momentum with the
Noether charges of global Poincar\'e transformations.  To express the angular
momentum solely in terms of an energy-momentum (pseudo) tensor, the
energy-momentum (pseudo) tensor needs to be symmetric under interchange of two
spacetime indices.  Our goal is thus to find an expression for the symmetric
energy-momentum (pseudo) tensor, which is conserved as in (1) and which is
given by the Noether procedure, rather than by manipulation of the field
equations of motion.

In the 2+1 dimensional Einstein gravity, asymptotically Minkowski boundary
conditions are not valid.$^1$ On the other hand, there is a gauge theoretical
formulation of the theory,$^2$ based on the Poincar\'e group $[ISO(2,1)]$. The
Noether charges associated with the Poincar\'e group gauge transformations
are identified as energy and angular momentum.

In 1+1 dimensions, we consider a gauge theoretical formulation$^{3}$ of the
string-inspired gravity model$^4$ and obtain an expression for energy arising
from the gauge transformations. Another way of finding an expression for
energy in 1+1 dimensions is to use the ADM definition;$^5$ we compare these
two approaches.

In Section~II, we analyze in a systematic way general properties of the
Noether charge associated with a local symmetry and also symmetrization of the
energy-momentum tensor (``improvement'').

In Section~III, the 3+1 dimensional Einstein-Hilbert action is investigated
and a symmetric energy-momentum (pseudo) tensor, as an improved Noether
current, is derived and compared with other definitions that have appeared in
the literature. Also we study the conserved Noether currents associated with
diffeomorphism invariance, both in the Einstein-Hilbert and the Palatini
(first-order) formulation.

Since asymptotically Minkowski boundary conditions can not be imposed in 2+1
dimensional Einstein gravity, we obtain in Section~IV expressions for energy
and angular momentum in the context of the gauge theoretical formulation for
the theory.

In Section~V, we consider a gauge theoretical formulation of 1+1 dimensional
gravity. After getting an expression for energy, we show that it agrees with
the ADM energy.

Concluding remarks comprise the final Section VI.

\goodbreak\bigskip
\line{\bf II.~~CONSERVATION LAWS \hfil}
\nobreak\medskip\nobreak\noindent
The Noether current associated with a local symmetry can always be brought to
a form that is identically conserved. This was shown by E.~Noether,$^6$ but
unlike the construction of conserved currents associated with a global
symmetry, her argument has not found its way into field theory textbooks ---
so we give a general proof in the Appendix.

To illustrate the result in a special example, let us consider the
Maxwell-scalar system, with a Lagrange density
$$
\cl = - {1\over4 }F_{\mu\nu}F^{\mu\nu}
+ (\!D^{\mu}\phi\!)^*D_\mu\phi
\eqno{(3)}
$$
where $D_\mu\phi \equiv (\partial_\mu + ieA_\mu)\phi$ and $F_{\mu\nu} \equiv
\partial_\mu A_\nu - \partial_\nu A_\mu $.
The Lagrangian $\cl$ is invariant under a local $U(1)$ gauge symmetry.
$$
\eqalign{
\phi'(x) &= e^{-\! ie\theta (x)} \phi(x),\ \delta \phi =- ie\theta\phi \cr
A'_{\mu}(x) &= A_\mu (x) + \partial_\mu \theta (x),
{}~~\delta A_\mu = \pp_\mu \theta \cr}
\eqno{(4)}
$$
The associated Noether current
$$
\eqalign{
j^\mu &= {\pp \cl \over \pp\pp_\mu {A}_\nu}\delta A_\nu +
{\pp \cl \over \pp\pp_\mu \phi}\delta \phi +
{\pp \cl \over \pp\pp_\mu \phi^*}\delta\phi^* \cr
&= -F^{\mu\nu}\pp_\nu\theta
- (D^\mu\!\phi)^*ie\theta\phi + D^\mu\!\phi ie\theta\phi^* \cr}
\eqno{(5)}
$$
can be written with use of the equation of motion
$$
\pp_\nu F^{\nu\mu} = 2e\,{\rm Im}\left[(D^\mu\!\phi)^*\phi\right]
\eqno{(6)}
$$
as
$$ j^\mu = \pp_\nu\left( F^{\nu\mu}\theta \right) \eqno{(7)} $$
which is certainly identically conserved, regardless whether $F^{\mu\nu}$
satisfies the field equations, since the quantity in the parenthesis of (7) is
antisymmetric under the interchange of the indices $\mu$ and $\nu$.

The Noether charge is constructed as a volume integral of the time component
$j^{0}$.
$$
Q = \int_V dV\pp_i\left[F^{i0}(x)\theta (x)\right]
  = \int_{\pp V}dS^i F^{i0}(x)\theta (x)
\eqno{(8)}
$$
Without suitable boundary conditions, this charge either diverges or vanishes,
and in general does not lead to a conserved quantity. Moreover, even if we get
a finite value for $Q$ with some $\theta (x)$, the time dependence of $Q$ is
completely determined by the specified boundary condition.

An example of boundary conditions for (7) is
$$
\eqalign{
F^{0i} &\sim o \, ({1\over r^2}) \cr
\pp_{0} F^{0i} &\sim o \, ({1\over r^3}) \cr}
{\hskip.8in}{\rm as~} r \rightarrow \infty
\eqno{(9)}
$$
The first condition gives finite $Q$ when $\theta$ is constant at infinity,
and the second condition ensures that $Q$ is time independent. The asymptotic
condition that $\theta$ be constant can be extended through all space, thereby
arriving at a Noether formula for the total charge arising from a global
transformation.

Next, let us review the symmetrization procedure of the energy-momentum tensor
which was originally presented by Belinfante$^7$,
and which is always available in a Poincar\'e invariant theory.
Here, we generalize his
method to the case that the Lagrangian contains second derivatives, as is true
of the Einstein-Hilbert action.

Thus, consider
$$
I = \int_\Omega dx \, \cl
\left(\phi, \pp_\mu \phi, \pp_\mu\pp_\nu \phi \right)
\eqno{(10)}
$$
where $\phi$ is a multiplet of fields, and suppose $I$ is invariant under
Poincar\'e transformations. Under the infinitesimal action of these
transformations, coordinates and fields transform respectively by
$$
\eqalign{
x^\mu &\rightarrow x'^\mu = x^\mu - \kappal^{\mu}_{~\nu} x^\nu -a^\mu \cr
\phi(x) &\rightarrow \phi' (x') = L \, \phi(x) \cr}
\eqno{(11)}
$$
where
$ L = 1 + {1\over 2} \kappal^{\mu}_{~\nu} S^{\nu}_{~\mu} $
is a representation of the Lorentz group and
$ \kappal^{\mu}_{~\nu},~~S^{\nu}_{~\mu} $ satisfy the following relations.
$$
\eqalign{
\kappal^0_{~0} &= ~S^0_{~0} = 0 \cr
\kappal^0_{~i} &= ~\kappal^i_{~0}, ~S^0_{~i} = ~S^i_{~0} \cr
\kappal^i_{~j} &= -\kappal^j_{~i}, ~S^i_{~j} = -S^j_{~i} \cr}
\eqno{(12)}
$$
To derive the Noether current, let us consider the variation of the Lagrange
density under the transformations (11),
$$
\delta\cl = {\pp\cl\over\pp\phi}\,\delta\phi
+ {\pp\cl\over\pp\pp_\mu\phi}\,\pp_\mu\delta\phi
+ {\pp\cl\over\pp\pp_\mu\pp_\nu\phi}\,\pp_\mu\pp_\nu\delta\phi
\eqno{(13)}
$$
where $\delta \phi$ denotes $\phi' (x) -\phi(x)$.
Since the action is Poincar\'e invariant by hypothesis, $\delta \cl$ can
be written as a total derivative without using the equations of motion.
$$
\eqalign{
\delta\cl &= \pp_\mu \left( f^\mu\cl \right) \cr
f^\mu &\equiv \kappal^\mu_{~\nu} \, x^\nu + a^\mu \cr}
\eqno{(14)}
$$
On the other hand, using the
Euler-Lagrange equation
$$
{\pp\cl\over \pp\phi} - \pp_\mu\,{\pp\cl\over\pp\pp_\mu\phi}
+ \pp_\mu\pp_\nu\,{\pp\cl\over\pp\pp_\mu\pp_\nu\phi}=0
\eqno{(15)}
$$
we can rewrite (13) as a total derivative.
$$
\delta \cl= \pp_\mu\left[
{\pp\cl\over \pp\pp_\mu\phi}\delta\phi +
{\pp\cl\over \pp\pp_\mu\pp_\nu\phi}\pp_\nu\delta\phi
-\pp_\nu{\pp\cl\over\pp\pp_\mu\pp_\nu\phi}\delta\phi\right]
\eqno{(16)}
$$
Equating the above two expressions for $\delta\cl$, (14) and (16),
we arrive at a conservation equation.
$$
\pp_\mu\left[-f^\mu\cl+
{\pp\cl\over \pp\pp_\mu\phi}\delta\phi +
{\pp\cl\over \pp\pp_\mu\pp_\nu\phi}\pp_\nu\delta\phi
-\pp_\nu{\pp\cl\over\pp\pp_\mu\pp_\nu\phi}\delta\phi\right]=0
\eqno{(17)}
$$
Inserting now the variation, see (11),
$$
\delta\phi = f^\mu\pp_\mu\phi+{1\over 2}\kappal^\mu_{~\nu} S^\nu_{~\mu}\,\phi
\eqno{(18)}
$$
into (17), we get
$$
\pp_\mu \left[ f^\alpha \TC{}^\mu_{~\alpha} + {1\over2}
\kappal^\alpha_{~\beta} \, L^{\mu\beta}{}_{\!\alpha} \right]=0
\eqno{(19)}
$$
where
$\TC{}^\mu_{~\alpha}$
is the unsymmetric, canonical energy-momentum tensor,
$$
\TC{}^\mu_{~\alpha}
= -\delta^\mu_\alpha\cl + {\pp\cl\over
\pp\pp_\mu\phi}\pp_\alpha\phi
+ {\pp\cl\over \pp\pp_\mu\pp_\nu\phi}\pp_\nu\pp_\alpha\phi
-\pp_\nu{\pp\cl\over \pp\pp_\mu\pp_\nu\phi}\pp_\alpha\phi
\eqno{(20)}
$$
and
$$
\eqalign{
L^{\mu 0}{}_{\!0} & = 0 \cr
L^{\mu i}{}_{\!0} & = L^{\mu 0}{}_{\!i} \cr
  & = {\pp\cl\over\pp\pp_\mu\phi} \, S^i_{~0} \, \phi
    + {\pp\cl\over\pp\pp_\mu\pp_\nu\phi} \, S^i_{~0} \, \pp_\nu\phi
    - \pp_\nu {\pp\cl\over\pp\pp_\mu\pp_\nu\phi} \, S^i_{~0} \phi
    + \left( {\pp\cl\over\pp\pp_\mu\pp_0\phi} \, \pp_i\phi
    + {\pp\cl\over\pp\pp_\mu\pp_i\phi} \, \pp_0\phi \right) \cr
L^{\mu i}{}_{\!j} & =
  {\pp\cl\over\pp\pp_\mu\phi} \, S^i_{~j} \phi
  + {\pp\cl\over\pp\pp_\mu\pp_\nu\phi} \, S^i_{~j} \pp_\nu\phi
  - \pp_\nu \, {\pp\cl\over\pp\pp_\mu\pp_\nu\phi} \, S^i_{~j} \phi
  - \left({\pp\cl\over \pp\pp_\mu\pp_j\phi} \, \pp_i\phi
  - {\pp\cl\over \pp\pp_\mu\pp_i\phi} \, \pp_j\phi \right) \cr}
\eqno{(21)}
$$
As it is seen, Lorentz invariance of the action needs no reference to a
background Minkowski metric. Nevertheless formulas (12) and (21) can be
presented compactly by moving indices with the help of the flat metric
$\eta_{\mu\nu} = {\rm diag}(1,-\!1,\cdots ,-\!1)$.
Thus with the definitions
$\kappal_{\alpha\beta}=\eta_{\alpha\mu} \kappal^{\mu}_{~\beta}$,
$S^{\alpha\beta}=S^{\alpha}_{~\mu} \eta^{\mu\beta}$
and
$ L^{\mu\alpha\beta} = L^{\mu\alpha}{}_{\!\nu} \eta^{\nu\beta}$,
we see that the newly defined quantities
are antisymmetric in $\alpha$ and $\beta$.

Next we define $h^{\mu\beta\alpha}$ as
$$
h^{\mu\beta\alpha} =
{1\over 2}
\left[
L^{\mu\beta\alpha}
+ L^{\alpha\beta\mu}
+ L^{\beta\alpha\mu}
\right]
\eqno{(22)}
$$
so that it is antisymmetric in $\mu$ and $\beta$ and
${1\over 2}\kappal_{\alpha\beta}L^{\mu\beta\alpha}$ is identical to
$\kappal_{\alpha\beta}h^{\mu\beta\alpha}$.
Using these properties of
$h^{\mu\beta\alpha}$ and
$\kappal_{\alpha\beta} = - \pp_\alpha f_{\beta}$
[$\beta$ is lowered with $\eta_{\alpha\beta}$],
we finally get
$$
\pp_\mu\left[ f_\alpha\left( \TC{}^{\mu\alpha} +
\pp_\nu h^{\mu\alpha\nu}\right)\right] = 0
\eqno{(23)}
$$
[$\alpha$ is raised with $\eta^{\alpha\beta}$.]
Upon taking $f_{\alpha} =a_\alpha$,
we arrive at the conserved energy-momentum tensor.
$$
\Theta^{\mu\nu} = \TC{}^{\mu\nu}+\pp_\alpha h^{\mu\nu\alpha}
\eqno{(24)}
$$
To prove that $\Theta^{\mu\nu}$ is symmetric,
take $f_{\alpha}$ to be $\kappal_{\alpha\beta} x^\beta$.
Since (23) holds for arbitrary antisymmetric $\kappal_{\alpha\beta}$,
it follows that
$$
\pp_\mu\left[x^\alpha \Theta^{\mu\nu} - x^\nu \Theta^{\mu\alpha}\right] = 0
\eqno{(25)}
$$
The conservation law $\pp_\mu\Theta^{\mu\nu}=0$ and (25) imply that
$\Theta^{\mu\nu}$ is symmetric.

In conclusion, we have derived an expression for a conserved and symmetric
energy-momentum tensor for a Poincar\'e invariant theory whose action may
contain second derivatives.

\goodbreak\bigskip
\line{\bf III.~~GRAVITATIONAL ENERGY-MOMENTUM (PSEUDO) TENSOR IN \hfil}
\nobreak
\line{\hskip.5in \bf 3+1 DIMENSIONS \hfil}
\nobreak\medskip\nobreak\noindent
The Einstein-Hilbert action
$$
I = -{1\over 16\pi k}\int d^4x \, \sqg R +I_M
\eqno{(26)}
$$
where $k$ is the gravitational coupling, $R$ the scalar curvature, and $I_M$
denotes a matter action, can be put into a form involving only first
derivatives through an integration-by-part of terms involving the second
derivatives.  The explicit form of the first-derivative action is
$$
\bar I = -{1\over16\pi k} \int d^4x \, \sqg G + I_M
\eqno{(27)}
$$
where $G$ is given in terms of the Christoffel connections $\Gamma$.
$$
G = g^{\mu\nu}
\left( \Gamma^\alpha_{\mu\beta} \Gamma^\beta_{\nu\alpha}-
       \Gamma^\alpha_{\mu\nu}   \Gamma^\beta_{\alpha\beta} \right)
\eqno{(28)}
$$
To be specific, let us take the matter action for a massless scalar.
$$
I_M = {1\over 2} \int d^4x \sqg g^{\mu\nu} \pp_\mu \phi\pp_\nu \phi
\eqno{(29)}
$$
The same equation of motion follows from $I$ and $\bar{I}$,
$$
8\pi k \, T_{\mu\nu} = R_{\mu\nu} - {1\over2} g_{\mu\nu} \, R
\eqno{(30)}
$$
where $T_{\mu\nu}$ is the matter energy-momentum tensor
$T_{\mu\nu} = {2 \over \sqrt{-g}} \, {\delta I_M \over \delta g^{\mu\nu}}$.

Although the action $\bar I$ is conventional in that only first derivatives
occur, its integrand $G$ is no longer a scalar. However, both $I$ and $\bar I$
are invariant under Poincar\'e transformations. Therefore we can use the
generalized Belinfante method to find an expression for the symmetric
energy-momentum (pseudo) tensor arising both from $I$ and $\bar I$.

Note that the spin matrix for the metric field $g_{\mu\nu}$ is given by
$$
\eqalign{
\left( S^{\alpha\beta}{g}\right)_{\mu\nu} &=
\left(\eta^{\beta\kappa}\delta^\alpha_\mu
- \eta^{\alpha\kappa}\delta^\beta_\mu\right)
g_{\kappa\nu}
+ \left(\eta^{\beta\kappa}\delta^\alpha_\nu
- \eta^{\alpha\kappa}\delta^\beta_\nu \right) g_{\mu\kappa}\cr}
\eqno{(31)}
$$

After some straightforward calculations following the generalized Belinfante
method, and with a bit more algebra using the equations of motion, we are led
to the following symmetric energy-momentum (pseudo) tensor from both $I$ and
$\bar I$.
$$
\Theta^{\mu\nu}
= {1\over16\pi k} \pp_\alpha \pp_\beta \sqg
\left[ \eta^{\mu\nu} g^{\alpha\beta} -\eta^{\alpha\nu} g^{\mu\beta}
     + \eta^{\alpha\beta} g^{\mu\nu} - \eta^{\mu\beta} g^{\alpha\nu} \right]
\eqno{(32)}
$$

Let us compare the above result to other formulas for the symmetric
energy-momentum (pseudo) tensor found in the literature. Although there are
many expressions for the gravitational energy and momentum,$^8$ there seem to
be only two for a {\it symmetric} energy-momentum (pseudo) tensor.  These are
obtained by manipulating the Einstein field equation. The first one is
discussed by Landau and Lifshitz$^9$
$$
\Theta'^{\mu\nu} =
{1\over16\pi k} \pp_\alpha \pp_\beta
\left[ (-\! g) (g^{\mu\nu} g^{\alpha\beta}
               -g^{\alpha\nu} g^{\mu\beta}) \right]
\eqno{(33)}
$$
and the other by Weinberg$^{10}$
$$
\Theta''^{\mu\nu}
= {1\over16\pi k} \pp_\alpha \pp_\beta
\left[ \eta^{\mu\nu} A^{\alpha\beta} - \eta^{\alpha\nu} A^{\mu\beta}
     + \eta^{\alpha\beta} A^{\mu\nu} - \eta^{\mu\beta} A^{\alpha\nu} \right]
\eqno{(34)}
$$
where $A^{\alpha\beta} =
-h^{\alpha\beta} + {1\over 2} \eta^{\alpha\beta} \, h^\gamma_{~\gamma}$,~
$g_{\mu\nu} = \eta_{\mu\nu}+h_{\mu\nu}$, and indices are raised and lowered
with the flat metric.

Although obtained by totally different methods,
$\Theta'$ and $\Theta''$ agree with $\Theta$ in (32) up to first order in $h$.
The difference between $\Theta'$ and $\Theta$ is
$$
\Theta'^{\mu\nu} - \Theta^{\mu\nu}
= {1\over16\pi k} \pp_\alpha \pp_\beta
\left[ l^{\mu\nu} l^{\alpha\beta}
     - l^{\mu\beta} l^{\alpha\nu} \right]
\sim o \, (h^2)
\eqno{(35)}
$$
where $l^{\mu\nu}=\sqg g^{\mu\nu}-\eta^{\mu\nu}$; while
the difference between $\Theta''$ and $\Theta$ is
$$
\Theta''^{\mu\nu} - \Theta^{\mu\nu} ={1\over16\pi k} \pp_\alpha \pp_\beta
\left[
\eta^{\mu\nu} B^{\alpha\beta} - \eta^{\alpha\nu} B^{\mu\beta}
+ \eta^{\alpha\beta} B^{\mu\nu} - \eta^{\mu\beta} B^{\alpha\nu}
\right]
\sim o \, (h^2)
\eqno{(36)}
$$
where
$B^{\alpha\beta} =
- h^{\alpha\beta}
+ {1\over 2} \eta^{\alpha\beta} \, h^\gamma_{~\gamma}
- \sqg \, g^{\alpha\beta}$.

The corresponding expression
for the energy derived from $\Theta$ is at order $h$
$$ E = \int d^3\!r ~\Theta^{00} =
{1\over 16\pi k }\int dS^i \left[\pp_i h_{jj} -\pp_j h_{ij}\right]+o(h^2)
\eqno{(37)}
$$
while the angular momentum reads
$$
\eqalign{
J_{ij} &= \int d^3\! r \left( x^i \Theta^{0j} - x^j \Theta^{0i}\right) \cr
&= {1\over 16 \pi k}\int dS^k \left[\left(x^i\pp_0h_{jk}-
x^i\pp_k h_{0j} + \delta_{ki}h_{0j}\right) -(i\leftrightarrow j)\right]+o(h^2)
\cr}
\eqno{(38)}
$$

We evaluate these expressions on a solution to the Einstein's
equation with a rotating point source --- the Kerr solution ---
whose line element has the following large $r$ asymptote.
$$
\eqalign{
ds^2 &= \left( 1-{2km\over r} + {\rm o}(r^{-2}) \right) dt^2
- \left( 4kJ\epsilon_{ij3}{x^j\over r^4} +{\rm o}(r^{-4})\right)dx^idt \cr
&- \left(1+{2km\over r} + {\rm o}(r^{-2})\right)dx^idx^i \cr}
\eqno{(39)}
$$
We find $E=m$ and
$J_{ij}=J \, \epsilon_{ij}$.
[$E$, $E'$ and $E''$
(similarly $J_{ij}$, $J'_{ij}$ and $J''_{ij}$)
could be different from one another,
if the order $h$ terms in (37) and (38) vanish
and the terms of order $h^2$ survive;
this of course does {\it not} happen for the Kerr solution.]

Next, observe that the action $I$ is diffeomorphism invariant-- a
symmetry certainly bigger than the (global) Poincar\'e symmetry.
We are naturally led to inquire what is the conserved current
associated with this diffeomorphism invariance.  {}From (17), we
can read off the expression for the Noether current associated
with the diffeomorphism $\delta x^\mu = -f^\mu (x)$, where $f^\mu$
is an arbitrary function of $x$.
$$
j_f^\mu
= -f^\mu\cl
+ {\pp\cl\over \pp\pp_\mu\phi}\delta\phi
+ {\pp\cl\over \pp\pp_\mu\pp_\nu\phi}\pp_\nu\delta\phi
- \pp_\nu{\pp\cl\over\pp\pp_\mu\pp_\nu\phi}\delta\phi
\eqno{(40)}
$$
Starting from the action $I$, after some straightforward calculations, we get
$$
{1\over \sqg} j_f^\mu = T^{\mu}_{~\nu} f^\nu + {1\over16\pi k}
\left[ f^\mu R - D_\nu\left(D^\mu f^\nu
+ D^\nu f^\mu -2g^{\mu\nu}D_\alpha f^\alpha\right) \right]
\eqno{(41)}
$$
Using the equation of motion (30)
and the relation $\left[D^\nu,D^\mu \right]f_\nu = R^{\mu\nu}f_\nu$, we get a
remarkably simple expression,
$$
{1\over\sqg} j_f^\mu =
{1\over16\pi k} D_\nu \left[ D^\mu f^\nu - D^\nu f^\mu \right]
\eqno{(42)}
$$
which was first given by Komar$^{8}$ \footnote{**}
{Komar's formula is actually twice of (42).  Presumably, he reached his
expression by guesswork, so he did not obtain the factor ${1\over
2}$, which comes from the normalization of the action. Later,
P. G. Bergmann$^{11}$ derived (42) with the correct factor.}  and
is extensively discussed in the literature.

In spite of the simple and appealing formula (42) for the current,
we encounter the following difficulty in attempting to use it in a
definition of energy.  For $f^\mu = \delta^\mu_0$,
$E_N=\int d^3\!r \, j_f^0$
gives only half of the expected energy for the Kerr
solution (39) [note that $E_N$ is not obtained from a symmetric
tensor while $E$ in (37) is].  But we can not simply
``renormalize'' $j_f^\mu$ by a factor of two and get universal
agreement with previous formulas. This is because if we construct
the angular momentum generator from (42)
$\int d^3\!r \, j^0_f,~~f^i=
\kappal^i_{~j} x^j$, the expression agrees with that from (38) at
order $h$, and gives the correct answer in the Kerr case.

The action in (26) may alternatively be presented in first-order,
Palatini form,
$$
\eqalign{
I = &{1\over 64\pi k} \int d^4\! x\epsilon^{\alpha\beta\gamma\delta}
\epsilon_{ABCD} e^A_\alpha e^B_\beta R^{CD}_{\gamma\delta} \cr
    & R^{CD}_{\gamma\delta}\equiv \pp_\gamma\omega^{CD}_\delta +
    \omega^{CE}_\gamma \omega_{\delta E}\!^D - (\gamma\leftrightarrow\delta)
\cr}
\eqno{(43)}
$$
where $e^A_\mu$ and $\omega^{AB}_\mu$ are the {\it Vierbein} and the spin
connection. We assume that the {\it Vierbein} is invertible, with inverse
$e^\mu_A$ and $e \equiv$det$e^A_\mu$ equals $\sqg$.

Since this action is invariant under Poincar\'e transformations, let us
again construct the symmetric energy-momentum (pseudo) tensor. Noting that the
spin matrices for $e^A_\mu$ and $\omega^{AB}_\mu$ are given by
$$
\eqalign{
\left( S^{\alpha\beta} e^{A} \right)_\mu &=
\left( \delta^\beta_\mu\eta^{\alpha\nu}
- \delta^\alpha_\mu\eta^{\beta\nu}\right)
e^{A}_\nu \cr
\left( S^{\alpha\beta} \omega^{AB}\right)_\mu &=
\left( \delta^\beta_\mu\eta^{\alpha\nu}
- \delta^\alpha_\mu\eta^{\beta\nu}\right)
\omega^{AB}_\nu \cr}
\eqno{(44)}
$$
and going through the generalized Belinfante procedure, one finds that the
symmetric tensor vanishes.  This is because the added superpotential, needed
to symmetrize the nonsymmetric, canonical
energy-momentum (pseudo) tensor, exactly cancels it.

The action in (43) is also diffeomorphism invariant, with the {\it Vierbein}
and the spin connection transforming as
$$
\eqalign{
\delta_f e^A_\mu &=f^\alpha\pp_\alpha e^A_\mu + \pp_\mu f^\alpha e^A_\alpha \cr
\delta_f \omega^{AB}_\mu &=f^\alpha\pp_\alpha \omega^{AB}_\mu+ \pp_\mu f^\alpha
\omega^{AB}_\alpha \cr}
\eqno{(45)}
$$
Let us compute the Noether current associated with diffeomorphism invariance of
the action (43), analogous to
the Komar current (42) for the action (26). The resulting expression is
$$
j_f^\mu
= -{1\over8\pi k} \pp_\nu
\left( e e^\mu_A e^\nu_B\omega^{AB}_\alpha f^\alpha \right)
\eqno{(46)}
$$
It is again identically conserved because the quantity in the parenthesis of
(46) is antisymmetric in $\mu$ and $\nu$.  Taking $f^\alpha = \delta_0^\alpha$
and integrating over the space, we get yet another expression for energy.
$$
E_P = -{1\over8\pi k} \int dS^i \, e e^0_A e^i_B \omega^{AB}_0
\eqno{(47)}
$$
Note that this is different from the Komar formula for $E_N$ from (42),
and from $E$ in (37),
even in asymptotic form.
We evaluate (47) for the Kerr line element
(39) and find $E_P=m/2$, as in the Komar expression.

Before discussing angular momentum, let us observe that the action in (43) is
also invariant under gauge transformations
$$
\eqalign{
\delta e^A_\mu &= \kappa^A\!_B e^B_\mu \cr
\delta \omega^{AB}_\mu &=\pp_\mu\kappa^{AB}+
\kappa^A\!_C\omega^{CB}_\mu-\kappa^B\!_C\omega^{AC}_\mu \cr}
\eqno{(48)}
$$
where $\kappa^{AB}$ is an arbitrary function of $x$ and antisymmetric in $A$
and $B$.  Since the gauge group associated with this symmetry is the Lorentz
group, its spatial generators are usually identified with angular
momentum. Now some straightforward calculations give us a Noether current
associated with this gauge transformation.
$$
j_\kappa^\mu
= -{1\over 8 \pi k} \pp_\nu
\left( e e^\mu_A e^\nu_B \kappa^{AB} \right)
\eqno{(49)}
$$
[Note that (46) follows from (49) when the gauge function $\kappa^{AB}$ is
taken as $\omega^{AB}_\mu f^\mu$.]  The corresponding expressions of angular
momentum for the currents in (46), with $f^i = \kappal^i\!\,_jx^j$ and (49)
with $\kappa^{AB} =
(\delta_{a}^{~A} \delta_{b}^{~B} - \delta_{b}^{~A} \delta_{a}^{~B})$,
are respectively
$$
\eqalignno{
J'_{ab} &=
-{1\over8\pi k} \int dS^i \,
e e^0_A e^i_B (\omega^{AB}_a x_b - \omega^{AB}_b x_a) &{(50)}\cr
J''_{ab} &=
-{1\over8\pi k} \int dS^i \,
e (e^0_a e^i_b - e^0_b e^i_a) &{(51)}\cr}
$$
where $a$ and $b$ range over $(1,2,3)$.
With the line element in (47),
$J'_{ab}$ and $J''_{ab}$ are evaluated to $-{2\over 5}J
(\delta_{1a}\delta_{2b}-\delta_{2a}\delta_{1b})$ and $-{2\over 3}J
(\delta_{1a}\delta_{2b}-\delta_{2a}\delta_{1b})$ respectively,
rather than the expected value
$J (\delta_{1a}\delta_{2b}-\delta_{2a}\delta_{1b})$,
which follows from (38).  We have no explanation for the peculiar numerical
factors.

Another attractive formula for a conserved spacetime current,
also due to Komar, is
$$
{1\over \sqg} j^\mu = T^\mu_{~\nu} k^\nu
\eqno{(52)}
$$
where $k^\nu$ is a Killing vector. Owing to the covariant conservation of the
matter energy-momentum tensor, $ j^\mu$ is conserved, and therefore for static
solutions, with $k^\nu$ the Killing vector $k^\nu = \delta^\nu_0$ one is
tempted to identify $\int d^3\! r \sqg T^0_{~0}$ as the energy. However, as
is well known, the total mass of a static radially symmetric solution to
Einstein's equations is $\int d^3\!r \, T^0_{~0}$, hence (52) does not
function properly as a spacetime current.

\goodbreak\bigskip
\line{\bf IV.~~ENERGY AND ANGULAR MOMENTUM IN 2+1 DIMENSIONAL \hfil}
\nobreak
\line{\hskip.5in \bf GRAVITY \hfil}
\nobreak\medskip\nobreak\noindent
We begin by considering the 2+1 dimensional Einstein-Hilbert action.
$$
I = -{1\over 16\pi k}\int d^3\!x \, \sg R
\eqno{(53)}
$$
Because the Einstein and curvature tensors are equivalent, spacetime is flat
outside sources.  Therefore, all effects of localized sources are on the
global geometry.  In the presence of such global effects, spacetime is not
asymptotically Minkowski.  For example, we can solve the Einstein equation
for a rotating point mass (string in 3+1 dimensions). The solution
is described by the line element$^1$
$$
ds^2 = (dt+4kJd\theta)^2 -dr^2-(1-4km)^2r^2d\theta^2
\eqno{(54)}
$$
and there is no coordinate choice in which the asymptote is
Minkowski spacetime. The best one can do is to make the line element
``locally Minkowski'',
$$
ds^2 = d\tau^2-{dx'}^2 -{dy'}^2
\eqno{(55)}
$$
through the redefinitions
$$
\tau=t+4kJ\theta,~~x'=r{\rm cos}(1-4km)\theta,~~y'=r{\rm sin}(1-4km)\theta
\eqno{(56)}
$$
However, the angular range of $(1-4km)\theta $ is diminished to $(1-4km)2\pi$,
while $\tau$ jumps by $8\pi kJ$ whenever the origin is circumnavigated. Such
geometry is conical and not globally Minkowski.

For another viewpoint, let us consider this theory as the Poincar\'e
$ISO(2,1)$ gauge theory of gravity.$^2$ Here, we can exploit the possibility
of relating charges associated with gauge transformations to energy and
angular momentum.

The commutation relations of the Poincar\'e $ISO(2,1)$ group are
$$
\eqalign{
[P_A,P_B] &= 0, ~~[J_A,J_B] = \epsilon_{AB}\!\!\ ^CJ_C \cr
[J_A,P_B] &= \epsilon_{AB}\!\!\ ^CP_C \cr}
\eqno{(57)}
$$
where indices are raised or lowered by $\eta_{AB}$, and $\epsilon^{012} = 1$.
In Poincar\'e invariant field theories, $P_A$'s are interpreted as translation
generators,
$J_0$ is interpreted as angular momentum generator and the two $J_i$'s as
boosts.

If we introduce a connection
one-form
$A=e^AP_A + \omega^AJ_A$, where $e^A$ and
$\omega^A$ are respectively the {\it Dreibein\/} and the spin connection,
the curvature two-form is given by
$$
\eqalign{
F &= dA+A^2 \cr
  &= (de^A + \epsilon^A_{~BC} \omega^Be^C)P_A
   + (d\omega^A+{1\over 2} \epsilon^A_{~BC} \omega^B \omega^C) J_A \cr}
\eqno{(58)}
$$

The Chern-Simons action for this connection is
$$
I = {1\over 16\pi k}\int \langle A ~,~ dA+{2\over 3}A^2 \rangle
\eqno{(59)}
$$
with $\langle ~,~ \rangle$ denoting an invariant bilinear form in the algebra.
$$
\langle J_A ~,~ P_B \rangle = \eta_{AB},
{}~~\langle P_A ~,~ P_B \rangle = \langle J_A ~,~ J_B \rangle=0
\eqno{(60)}
$$
One verifies that (59) is a first order Palatini action equivalent to (53).

The generator of gauge transformations is also an element of the algebra:
$\theta =\alpha^AP_A+\beta^AJ_A$ with $\alpha^A$ and $\beta^A$ being
infinitesimal parameters. The variation of $A$ under a gauge transformation is
$$
\delta A = d\theta + [A,\theta]
\eqno{(61)}
$$
Note that the Lagrange density in the action
changes under the gauge transformation by a total derivative.
$$
\delta\cl = \pp_\mu X^\mu
\eqno{(62)}
$$
where
$X^\mu={1\over 16\pi k}\epsilon^{\mu\nu\rho}
\langle A_\nu ~,~ \pp_\rho\theta \rangle $.
Therefore the Noether current associated with this gauge transformation is
$$
j^\mu
= \langle {\pp\cl\over \pp\pp_\mu A_\nu} ~,~ \delta A_\nu \rangle - X^\mu
\eqno{(63)}
$$
Using the equation of motion $(F=0)$, we get
$$
j^\mu
= {1\over8\pi k} \epsilon^{\mu\nu\rho} \pp_\nu
\langle A_\rho ~,~ \theta \rangle
\eqno{(64)}
$$
which is an identically conserved current as expected and totally dependent on
the choice of gauge function $\theta $.

The solution to $F=0$, which leads to (54), gives rise to
the following {\it Dreibein} and spin connection$^{12}$
$$
\eqalign{
e^0 &= dt+{4 kJ\over r^2}{\bf r} \times d{\bf r} \cr
\pmb{e} ~&=(1-4km)d{\bf r}+{4km\over r^2}{\bf r}
\left({\bf r}\cdot d{\bf r} \right)\cr
\omega^0&={4km\over r^2}{\bf r} \times d{\bf r}\cr
\pmb{\omega} ~&=0\cr}
\eqno{(65)}
$$
with $x^1=r{\rm cos}\theta$ and $x^2=r{\rm sin}\theta$.
We inquire if ``charges'' coming from (63) and (64) could be identified
as energy and angular momentum, with values $m$ and $J$ respectively on the
solution (65).

To proceed we must choose a ``global'' form for $\theta$ in (64). A natural
choice is to take $\theta$ to be a constant along the $P^0$ direction for
defining energy and another constant along the $J^0$ direction for defining
angular momentum.  With this one finds
$$
E \equiv {1\over8\pi k} \oint dx^i \omega^0_i = m \,,
{}~~~ J \equiv {1\over8\pi k} \oint dx^i e^0_i = J
\eqno{(66)}
$$

Another choice for $\theta$ could be the following. We recall the relation
between a diffeomorphism implemented by a Lie derivative on a gauge potential
(connection) and a gauge transformation.$^{13}$
$$
\eqalign{
\delta_f A_\mu =L_f A_\mu &=f^\alpha\pp_\alpha A_\mu
+\pp_\mu f^\alpha A_\alpha \cr
&=f^\alpha F_{\alpha\mu}+\pp_\mu\left(f^\alpha A_\alpha\right)+
\left[A_\mu , f^\alpha A_\alpha\right] \cr}
\eqno{(67)}
$$
In this theory $F_{\alpha\mu}$ vanishes on shell, and it is natural
to identify the gauge transformation generated by $f^\alpha A_\alpha$ with the
infinitesimal diffeomorphism $f^\alpha$. With this choice, we find
$$
Q_f={1\over 8\pi k}\int dx^i
\langle A_i ~,~ A_\alpha f^\alpha \rangle
\eqno{(68)}
$$
which with (65), becomes
$$
Q_f = {1\over8\pi k} \int dx^i
\left[
\omega^0_i f^0 +\left( e^0_i \omega^0_j + \omega^0_i e^0_j \right) f^j
\right]
\eqno{(69)}
$$
For energy we take $f^0=1$ and $f^i = 0$, thereby again one finds
$$
E=m
\eqno{(70)}
$$
However for angular momentum,
where $f^0=0$ and $f^i=\epsilon^{ij}x^j$,
one gets $8kmJ$.
We do not have an explanation for the dimensionless factor $8km$.

\goodbreak\bigskip
\line{\bf V.~~1+1 DIMENSIONAL ENERGY IN GAUGE THEORETICAL \hfil}
\nobreak
\line{\hskip.5in \bf FORMULATION \hfil}
\nobreak\medskip\nobreak\noindent
In 1+1 dimensions,
the action of string-inspired gravity theory$^4$
can be written as
$$
I_g = \int d^2x\sqg \left( \eta R - \Lambda \right)
\eqno{(71)}
$$
where the ``physical'' metric
${\bar g}_{\mu\nu}$
is $g_{\mu\nu}/\eta$
and $R$ is the scalar curvature constructed from $g_{\mu\nu}$.

This theory is reformulated as a gauge theory
using a centrally extended Poincar\'e group,$^3$ whose algebra is
$$
\left[P_a,J\right]
=\epsilon_a\!\!\ ^b P_b,\ \ \ \left[P_a,P_b\right]= \epsilon_{ab}I
\eqno{(72)}
$$
The connection one-form $A$ and the curvature two-form $F$ are explicitly
$$
\eqalign{
A &= e^aP_a + \omega J + a I \cr
F &= dA+ A^2 = f^aP_a +f J + g I \cr
  &= \left( De \right)^a P_a + d\omega J
   + \left( da +{1\over 2} e^a \epsilon_{ab} e^b \right) I \cr}
\eqno{(73)}
$$
where $e^a$ and $\omega$ are the {\it Zweibein}
and the spin connection respectively.
The action
$$
\eqalign{
I &= \int \sum^3_{A=0} \eta_A F^A = \int \left[ \eta_a \left(De\right)^a +
\eta_2 d\omega+ \eta_3\left( da + {1\over 2}e^a\epsilon_{ab} e^b\right) \right]
\cr
&F^A = (f^a,f,g),
{}~~~\eta_A = (\eta_a, \eta_2,\eta_3),
{}~~~\eta_2 = -\eta \cr}
\eqno{(74)}
$$
is equivalent to $I_g$ and
is invariant under gauge transformations,
$$
\eqalign{
A &\rightarrow U^{-1}AU + U^{-1}dU \cr
H &\rightarrow U^{-1}HU \cr}
\eqno{(75)}
$$
Here $H=\eta_a P^a - \eta_3 J-\eta_2 I$,
and $U$ is the gauge function
$e^{\theta^a P_a} e^ {\alpha J} e^{\beta I}$
with arbitrary local parameters
$\theta^A=(\theta^a, \alpha, \beta)$.
Note that the infinitesimal form of a gauge transformation on $A$ is explicitly
$$
\eqalign{
\delta e^a &= -\epsilon^a_{~b} \alpha e^b +
\epsilon^a_{~b} \theta^b\omega + d\theta^a \cr
\delta \omega&= d\alpha \cr
\delta a &= - \theta^a \epsilon_{ab} e^b + d\beta\cr}
\eqno{(76)}
$$
The Noether current associated to this gauge transformation is
$$
\eqalign{
j^\mu
&= {\pp\cl \over \pp\pp_\mu e^a_\nu}  \, \delta e^a_\nu
 + {\pp\cl\over \pp\pp_\mu\omega_\nu} \, \delta\omega_\nu
 + {\pp\cl\over\pp\pp_\mu a_\nu}      \, \delta a_\nu \cr
&= {\pp\cl\over\pp\pp_\mu A^A_\nu}    \, \delta A^A_\nu \cr}
\eqno{(77)}
$$
Using the equations of motion and (76), we get
$$
j^\mu
= \epsilon^{\mu\nu} \pp_\nu
\left(\eta_a\theta^a + \eta_2\alpha +\eta_3\beta \right)
= \epsilon^{\mu\nu}
\pp_\nu \left( \eta_A\theta^A \right)
\eqno{(78)}
$$

As anticipated, the current $j^\mu$ is identically conserved and again totally
dependent on the choice of gauge functions.

Infinitesimal diffeomorphisms are performed on shell by a gauge transformation
with gauge function $f^\alpha A_\alpha$ [cf. (67)].
$$
\eqalign{
\delta_f A_\mu &=
\pp_\mu (f^\alpha A_\alpha) + \left[A_\mu , f^\alpha A_\alpha\right]\cr
\delta_f H& = \left[H, f^\alpha A_\alpha\right]\cr}
\eqno{(79)}
$$

Therefore we get an expression for energy by taking in (78) $\theta =f^\alpha
A_\alpha$ and $f^\mu = (1,0)$.
$$
j_E^\mu = \epsilon^{\mu\nu} \pp_\nu
\left(\eta_a e_0^a + \eta_2\omega_0 + \eta_3 a_0 \right)
\eqno{(80)}
$$
$$
E = \left( \eta_a e_0^a + \eta_2 \omega_0 +\eta_3 a_0 \right)
{\Bigr|}^{+\!\infty}_{-\!\infty}
\eqno{(81)}
$$
\def\lll{{\lambda\sigma}}
To see what happens with an explicit solution,
let us consider the black hole solution
in presence of infalling matter, i.e. with
$T_{+\! +} = m\delta (\sigma +\tau )$.\footnote{***}
{See for example, Bilal and Kogan in Ref.~[5].}
$$
e^a_\mu =e^{\lambda\sigma}\delta^a_\mu,
{}~~~a_0= -{e^{2\lambda\sigma}\over 2\lambda},
{}~~~a_1 =0,
{}~~~\omega_0 = -\lambda,
{}~~~\omega_1 = 0
\eqno{(82.a)}
$$
$$
\eqalign{
\eta_2 &= -\left[e^{2\lll} +{m\over\lambda}\left(1-e^{\lambda (\sigma +\tau)}
\right)\theta (\sigma +\tau )\right] \cr
\eta_0 &= -2\lambda e^\lll + me^\lll \theta  (\sigma +\tau )\cr
\eta_1 &= me^\lll \theta (\sigma + \tau )\cr
\eta_3 &= -2\lambda^2\ \ \ \ \ \ (4\lambda^2 = \Lambda)\cr }
\eqno{(82.b)}
$$
Inserting the above solution into (81),
we get the expected value for the energy.
$$
E=m
\eqno{(83)}
$$

Let us compare this result with the ADM definition of energy.  To get an ADM
energy, we rewrite the Lagrange density for the gravity sector as
$$
\eqalign
{\cl &= \eta_a {\dot e}^a_1 + \eta_2 {\dot \omega}_1 + \eta_3 {\dot a}_1
+ e^a_0\left(\eta_3 \epsilon_{ab} e^b_1
+ \omega_1 \epsilon_a^b \eta_b + \eta_a'\right)\cr
&~~~+~ \omega_0\left(\epsilon^a_b \eta_a e^b_1 + \eta_2'\right) + a_0 \eta_3'
-\left(\eta_a e^a_0 + \eta_2\omega_0 + \eta_3 a_0\right)'\cr }
\eqno{(84)}
$$
where prime and dot denote derivatives on $\sigma$ and $\tau$ respectively.

Note that the variation of $\cl$ is
$$
\delta\cl = {\delta\cl\over\delta\phi} \delta\phi -
\left(\delta\eta_a e^a_0 + \delta\eta_2 \omega_0 + \delta \eta_3 a_0\right)'
\eqno{(85)}
$$
where $\phi$ denotes all the fields.
To eliminate boundary contributions, we have to introduce in the action
an appropriate boundary term and a boundary condition such that
boundary variations cancel.
The required boundary term is then
identified as the energy. Let us take the boundary condition to be
$$
A_0 \rightarrow  A_0\Big|_{{\rm free}} ~~~~as~~\sigma\rightarrow\pm\infty
\eqno{(86)}
$$
where subscript `free' denotes an empty space solution [(82.a)].
The necessary boundary
contributes to the Lagrange density a total derivative.
$$
\cl_B = \left( \eta_a e^a_0 \Big|_{{\rm free}}
             + \eta_2 \omega_0 \Big|_{{\rm free}}
             + \eta_3 a_0 \Big|_{{\rm free}} \right)'
\eqno{(87)}
$$
Therefore the  ADM energy is
$$
E_{ADM} = \left( \eta_a e^a \Big|_{{\rm free}}
               + \eta_2 \omega_0 \Big|_{{\rm free}}
               + \eta_3 a_0 \Big|_{{\rm free}} \right)
          \bigg|^{+\!\infty}_{-\!\infty}
\eqno{(88)}
$$
which coincides with the expression (81) with the help of (86).
Thus $E$ agrees with $E_{ADM}$.

We remark that without matter there is no step function in the solution (82.b)
and expressions (81) and (88) vanish, thus giving no mass to the `pure' black
hole configuration contrary to what is usually argued.\footnote{****}
{See for example, de Alwis in Ref.~[5].}

Finally note that if one takes $\theta$ to be a constant along the $P^0$
direction for defining energy [as we did in the 2+1 dimensional gauge gravity
theory] and calculates the energy using the solution (82), one gets a
diverging result.

\goodbreak\bigskip
{\noindent\bf VI.~~DISCUSSION}
\nobreak\medskip\nobreak\noindent
Noether's procedure for constructing conserved symmetry currents provides a
universal method, which in particular may be used to derive energy-momentum
(pseudo) tensors in various gravity theories. This allows for an {\it a
priori}, symmetry-motivated approach to the problem, in contrast to the
conventional construction, which relies on manipulating equations of motion,
and which is motivated {\it a posteriori}, even while a variety of results
emerges, reflecting the variety of possible manipulations on the equations of
motion.$^{9,10}$

However, the Noether procedure is not without ambiguity. Since a local symmetry
is operating, the symmetry current is a divergence of an antisymmetric
tensor.$^6$~
But the Noether method, which requires recognizing that the symmetry
variation of a Lagrange density is a total derivative: $ \delta \cl =\pp_\mu
X^\mu$, leaves an undetermined contribution to $X^\mu$, which also is the
divergence of an antisymmetric tensor. Moreover, as we have seen, a variety of
conserved currents may be derived, depending on whether one uses
Einstein-Hilbert or Palatini formulations, whether the coordinate invariance is
viewed as diffeomorphisms of geometrical variables or a gauge transformations
on gauge connections.  Finally observe that our expressions are neither
diffeomorphism nor gauge invariant. At the same time, in all instances one
Noether tensor gives the ``correct'' integrated expressions.

\vfill\eject
\centerline{\bf APPENDIX}
The Noether current associated with a local symmetry can always be brought in a
form that is identically conserved. This theorem is proved in Ref.~[6], but the
 form of the current in terms of the Lagrange density is not given.
Here we present the current explicitly for the case that the Lagrange density
contains at most first derivatives of fields and the
symmetry
variation of the field
does not depend on second or higher
derivatives of the parameter functions.

\def\ttt{\theta^A}
\def\ddd{\Delta_A}

Let us consider an action for a field multiplet $\phi$,
$$
I= \int_\Omega dx\cl (\phi,\pp\phi )
\eqno{(A1)}
$$
which is invariant under a local transformation,
$$
\delta \phi =\Delta_A \theta^A + \Delta^\mu_A\pp_\mu \theta^A
\eqno{(A2)}
$$
where $\ddd$ may depend on $\phi$ and  $\ttt$ is a gauge parameter function.
First we note the Noether identity
$$
{\delta I\over \delta\phi} \ddd -
\pp_\mu\left( {\delta I\over \delta\phi} \ddd^\mu\right)= 0
\eqno{(A3)}
$$
where ${\delta I\over \delta\phi}={\pp\cl\over\pp\phi}-
\pp_\alpha\left({\pp\cl\over\pp\pp_\alpha\phi}\right)$.
Eq.~(A3) is a consequence of the invariance of the action against the
transformation (A2) with arbitrary $\ttt$.
$\delta \cl$ for arbitrary $\ttt$ reads
$$
\delta\cl =M_A\ttt +M_A^\mu\pp_\mu\ttt+M_A^{\mu\nu}\pp_\mu\pp_\nu\ttt
\eqno{(A4)}
$$
where
$$
\eqalign{M_A\,\,&= {\pp\cl\over\pp\phi}\ddd +{\pp\cl\over\pp\pp_\alpha\phi}
\pp_\alpha\ddd \cr
M_A^\mu\,\,&= {\pp\cl\over\pp\phi}\ddd^\mu +{\pp\cl\over\pp\pp_\alpha\phi}
\left( \pp_\alpha\ddd^\mu + \delta_\alpha^{~\mu} \ddd \right) \cr
M_A^{\mu\nu}&= {\pp\cl\over\pp\pp_\mu\phi}\ddd^\nu
\equiv M^{\left(\mu\nu\right)}_A -{\cal F}^{\mu\nu}_A\cr}
\eqno{(A5)}
$$
In the last equation $M_A^{\mu\nu}$ is decomposed into its symmetric
[$M^{\left(\mu\nu\right)}_A$]
and antisymmetric [$ -{\cal F}^{\mu\nu}_A$] parts.
Using the Noether identity, one easily finds the relation,
$$
M_A=\pp_\mu M_A^\mu -\pp_\mu\pp_\nu M^{\left(\mu\nu\right)}_A
\eqno{(A6)}
$$
which shows that $\delta \cl$
can be presented as a total derivative, since the
transformation (A2) {\it is\/} a symmetry.
$$
\eqalign{
\delta\cl &=\pp_\mu (M_A^\mu\ttt) +
M_A^{\left( \mu\nu \right)}\pp_\mu\pp_\nu\ttt
- \pp_\mu \pp_\nu M_A^{(\mu\nu)} \, \ttt \cr
&=\pp_\mu\left[M_A^\mu\ttt +
M_A^{(\mu\nu)}\pp_\nu\ttt - \pp_\nu \, M_A^{(\mu\nu)} \, \ttt \right] \cr}
\eqno{(A7)}
$$

Therefore, the Noether current is
$$
j^\mu={\pp\cl\over\pp\pp_\mu\phi}\delta\phi -\left[M_A^\mu\ttt +
M_A^{(\mu\nu)}\pp_\nu\ttt - \pp_\nu \, M_A^{(\mu\nu)} \, \ttt\right]
\eqno{(A8)}
$$
Inserting $\delta\phi$ in (A2) into (A8), and after a little algebra,
wherein use is made of the equation of motion
${\delta I \over \delta \phi} = 0$,
we get the desired expression for $j^\mu$.
$$
j^\mu=\pp_\nu\left({\cal F}_A^{\nu\mu}\ttt\right)
\eqno{(A9)}
$$
For the Maxwell Lagrangian, (A9) reproduces (7).

\vfill\eject
\centerline{\bf REFERENCES}
\medskip
\item{1.}
S.~Deser, R.~Jackiw and G.~'t Hooft, {\it Ann.~Phys.\/} (NY)
{\bf 152} (1984) 220;
S.~Deser and R.~Jackiw, {\it Ann.~Phys.\/} (NY) {\bf 153} (1984) 405.
\medskip
\item{2.}
A.~Ach\'ucarro and P. Townsend, {\it Phys. Lett. B \/}
{\bf 180} (1986) 89; E. Witten, {\it Nucl. Phys. \/} {\bf B 311} (1988/89) 46.
\medskip
\item{3.}
D.~Cangemi and R. Jackiw, {\it Phys. Rev. Lett. \/} {\bf 69} (1992) 233.
\medskip
\item{4.}
C.~Callan, S.~Giddings, J.~A.~Harvey
and A.~Strominger, {\it Phys. Rev. D \/} {\bf 45} (1992) 1005;
H.~Verlinde,
in {\it Marcel Grossmann Meeting on General Relativity\/},
H.~Sato and T.~Nakamura, eds.,
(World Scientific, Singapore, 1992)).
\medskip
\item{5.}R. Arnowitt, S. Deser and C. W. Misner, {\it Phys. Rev.\/}
{\bf 116} (1959) 1322 and {\it Phys. Rev.\/}
{\bf 122} (1961) 997; A. Bilal and I. I. Kogan,
{\it Phys.~Rev.~D\/}~{\bf 47} (1993) 5408;
S.~P.~de Alwis, COLO-HEP-309 (August 1993).
\medskip
\item{6.}E. Noether, {\it Nachr. Ges. Wiss. G\"ottingen\/} {\bf 2} (1918) 235.
For a review, see J. G. Flecher, {\it Rev. Mod. Phys.\/} {\bf 32} (1960) 65.
\medskip
\item{7.}F.~J. Belinfante, {\it Physica\/}~{\bf VII} (1940) 449.
\medskip
\item{8.}A. Einstein,
{\it Berlin Ber. \/}(1915) 778; C. M{\o}ller, {\it Ann. Phys.\/}  (NY)
{\bf 4} (1958) 347; A.~Komar, {\it Phys. Rev. \/}{\bf 113} (1959) 934;
R.~D. Sorkin, {\it Contemporary Mathematics \/} {\bf 71} (1988) 23.
\medskip
\item{9.}
L.~D. Landau and E.~M. Lifshitz, {\it The Classical Theory of Fields \/}
(Pergamon, New York, 1980) p.~280.
\medskip
\item{10.}S. Weinberg,
{\it Gravitation and Cosmology \/} (John Wiley {\&} Sons, New York,
1972) p.~165.
\medskip
\item{11.}P. G. Bergmann, in
{\it Encyclopedia of Physics} {\bf IV}
(Springer-Verlag, Berlin, 1962) p. 238.
\medskip
\item{12.}P. de Sousa Gerbert, {\it Nucl. Phys.\/} {\bf B 346}
(1990) 440.
\medskip
\item{13.}R. Jackiw, {\it Phys. Rev. Lett. \/} {\bf 41}
(1978) 1635.
\medskip
\item{14.}D. Cangemi and R. Jackiw, {\it Phys. Lett. B \/} {\bf 299} (1993) 24.
\par
\vfill
\end